%% file: main.tex
\documentclass[10pt,letterpaper]{article}

\usepackage[margin=0.76in]{geometry}
\usepackage[T1]{fontenc}
\usepackage{newtxtext,newtxmath}
\usepackage{microtype}
\usepackage{xcolor}
\usepackage{graphicx}
\usepackage{booktabs}
\usepackage{array}
\usepackage{enumitem}
\usepackage{amsmath}
\usepackage[font=small,labelfont=bf]{caption}
\usepackage[compact]{titlesec}
\usepackage{xurl}
\usepackage[hidelinks]{hyperref}

\definecolor{linkblue}{HTML}{225E91}
\hypersetup{
  colorlinks=true,
  linkcolor=linkblue,
  citecolor=linkblue,
  urlcolor=linkblue,
  pdfauthor={Zhenyu Zhao and Roy Zhao},
  pdftitle={Runtime-Independent Persistent Agents: Preserving Identity, Memory, and Code Across Models, Harnesses, and Servers}
}

\setlist{nosep,leftmargin=*}
\titlespacing*{\section}{0pt}{1.5ex plus .4ex minus .2ex}{.65ex}
\titlespacing*{\subsection}{0pt}{1.15ex plus .3ex minus .2ex}{.4ex}
\newcommand{\agentcore}{\mathcal{P}}
\newcommand{\executionenv}{\mathcal{E}}
\newcommand{\interactions}{\mathcal{S}}
\newcommand{\deployed}{\mathcal{A}}

\begin{document}

\begin{center}
{\LARGE\bfseries Runtime-Independent Persistent Agents:\par}
\vspace{0.12em}
{\Large\bfseries Preserving Identity, Memory, and Code Across Models, Harnesses, and Servers\par}
\vspace{0.5em}
{\large Zhenyu Zhao$^{1}$ \qquad Roy Zhao$^{2}$\par}
\vspace{0.15em}
{\small $^{1}$Independent Researcher\par}
{\small $^{2}$Paul G. Allen School of Computer Science \& Engineering, University of Washington\par}
{\scriptsize Reference implementation:
\href{https://github.com/our-ark/enoch}{github.com/our-ark/enoch}\par}
\end{center}
\vspace{0.15em}

\begin{abstract}
Long-lived AI agents may replace their models, orchestration harnesses,
interaction surfaces and hosts while retaining identity, memory and executable
body lineage. We present a runtime-independent architecture that separates a
continuity-bearing substrate $\agentcore_t=(I_t,M_t,B_t)$ from replaceable
execution and interaction bindings. Six continuity invariants and a
quiesce--checkpoint--validate--bind--rehydrate--resume protocol govern state
preservation, capability changes and continuation authority. Enoch implements
the design through a reusable software body, private installed state, provider
contracts and fenced migration.

We report complementary evidence from frozen implementation snapshots and
three kinds of live study. An operational case retains an established
instance's identity, nonempty memory, body revision and historical task IDs
across host, joint Codex-to-Claude model/harness, and chat-surface substitutions.
A separate cross-host task resumes from a verified artifact checkpoint.
A supervised Codex--Muse--Codex study completes five ordinary round trips and
five matched same-runtime controls on their first attempts after a separate
qualification run. A planned worker interruption recovers through the native
task API with the checkpoint preserved, a stale reply rejected and one task
completion recorded. These results support mechanical continuity for the
tested deployments and bounded workflows. They do not establish behavioral
equivalence, arbitrary-task portability, general exactly-once external effects
or unattended reliability.
\end{abstract}

\section{Introduction}

An agent can outlive a chat. It can also outlive the model, harness, process,
machine, or provider that happens to execute its next action. A personal agent
may begin in one chat application, move from a laptop to a server, replace a
reasoning model, resume an unfinished task through another harness, and still be
treated by its user as the same agent. Existing systems contain many of the
necessary mechanisms -- persistent memory, resumable sessions, provider
adapters, durable queues, and versioned code -- but these mechanisms do not by
themselves define the agent boundary.

A common shorthand identifies an agent with a model plus an agent harness. It
captures a useful synchronic question: what currently generates behavior? It
does not answer the diachronic question: what must remain continuous for two
executions at different times to be executions of the same agent? If the model
is replaced, did the agent migrate or die? If the same memory is copied into two
processes, are both the original? If a chat session is reset, has a new agent
been created? These are architecture and lifecycle questions, not only prompt
engineering questions.

We propose a sharper boundary. The persistent agent is the continuity-bearing
substrate of identity, memory, and executable body. Models, harnesses, and
servers form a replaceable execution substrate. Chats, APIs, email, and user
interfaces are replaceable interaction surfaces. The model and harness remain
causally important: they shape current behavior and capability. A communication
surface determines where an interaction occurs. Neither is a necessary
invariant of longitudinal identity.

A chat session is one interaction binding. Identity, memory and unfinished
work persist when the body commits them to durable state; the provider-native
transcript alone is not the agent's identity-bearing core. The body supplies
communication capabilities through provider contracts, while the current
account, channel and session remain external bindings.

This paper extends the agent-owned software-body boundary introduced in prior
work~\cite{zhao2026body}. That work asks what users can possess, inspect,
evolve, and use for descent. The present paper asks a different systems
question: how the whole continuity-bearing substrate can be rebound to a new
deployment without silently creating a new agent, resetting its
history, or allowing two hosts to act as the sole authority.

RIPA (Runtime-Independent Persistent Agents) contributes a lifecycle contract for the longitudinal AI agent:
what must remain continuous, which deployment bindings may change, and who
may authorize the successor. Prior work provides logical-entity persistence,
verified memory transfer and authorized state
activation~\cite{bernstein2014orleans,ravindran2026portable,he2026continuity}.
RIPA ties these foundations to an explicit $(I,M,B)$ agent boundary, six
continuity invariants and a migration protocol spanning reasoner, harness,
host and interaction changes. The contribution lies in this agent-level
contract and its realization across deployments.

The paper makes three contributions:

\begin{enumerate}
  \item \textbf{A runtime-independent agent boundary.} We separate a logical
  persistent substrate $(I,M,B)$ from a replaceable execution substrate
  containing the reasoner, harness, and host, plus independently replaceable
  interaction surfaces.
  \item \textbf{Migration semantics and continuity invariants.} We distinguish
  migration from restart, evolution, identity update, replication, and descent,
  and define an authorized migration protocol with explicit failure semantics.
  \item \textbf{An implementation and bounded empirical evaluation.} Enoch's
  provider contracts and durable state realize the architecture. We report
  operational host, Claude and chat-surface substitutions, cross-host task
  continuation, and supervised Muse round trips with controls and a planned
  interruption, distinguishing each result's evidence and limits.
\end{enumerate}

We use \emph{identity} functionally, without claims of consciousness,
personhood or metaphysical persistence. \emph{Authorized system continuity}
means that lineage, state, body and execution authority satisfy the
architectural invariants. \emph{Behavioral identity fidelity} concerns how an
execution recalls, composes and enacts its installed identity, a complementary
question operationalized by PAI-Bench~\cite{zhao2026pai}. Unless qualified,
continuity in this paper denotes the former.

\section{System Model}

\subsection{Persistent substrate, execution substrate, and interaction surfaces}

At time $t$, define the continuity-bearing persistent substrate as

\begin{equation}
  \agentcore_t=(I_t,M_t,B_t),
\end{equation}

where $I_t$ is the agent's architectural identity representation, $M_t$ is
private durable memory and continuity-relevant workflow state, and $B_t$ is the
versioned executable body: code, prompts, tools, policies, tests, provider
contracts, and evolution mechanisms. $\agentcore_t$ carries the persistent
agent state; valid longitudinal continuity additionally requires attributable
lineage and transition authority over successive versions of $\agentcore$.
These continuity conditions are distinct from $I_t$, even if represented in colocated
metadata; the components of $\agentcore_t$ likewise remain logically distinct
when packaged together.

The replaceable execution substrate is

\begin{equation}
  \executionenv_t=(R_t,H_t,D_t),
\end{equation}

where $R_t$ is a language model or other reasoner, $H_t$ is the orchestration
harness that supplies inference and tool execution, and $D_t$ is the host
device, service manager, and available environment capabilities. Separately,
let

\begin{equation}
  \interactions_t=\{s_{t,1},\ldots,s_{t,n}\}
\end{equation}

denote the currently bound interaction surfaces: chat accounts and sessions,
APIs, email, graphical interfaces, or other endpoints through which the agent
and its environment exchange events. A deployed execution is

\begin{equation}
  \deployed_t=\agentcore_t\triangleright
  (\executionenv_t,\interactions_t).
\end{equation}

\begin{figure}[t]
  \centering
  \includegraphics[width=\linewidth]{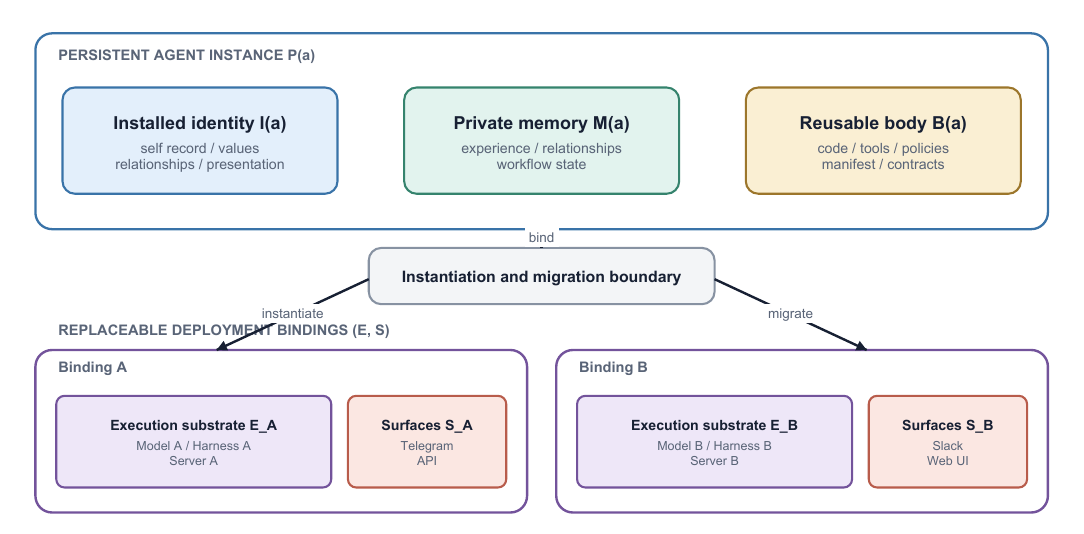}
  \caption{Runtime-independent agent boundary. An installed agent binds
  identity and private memory to an authorized reusable-body revision; each
  deployment supplies replaceable execution and interaction bindings.}
  \label{fig:runtime-boundary}
\end{figure}

The operator $\triangleright$ denotes instantiation, not ownership. The
execution substrate runs the persistent agent, while interaction surfaces
connect it to users and external systems; neither becomes its durable identity.
The body's communication interfaces belong to $B_t$, but a particular Slack
thread or API endpoint belongs to $\interactions_t$. Figure~\ref{fig:runtime-boundary}
shows two deployment bindings attached to the same logical substrate at
different times.

\subsection{Continuity is not bitwise immutability}

Persistence does not require all three components to remain byte-identical.
Memory grows, body code evolves through reviewed revisions, and identity may
change through an authorized governance process. The relevant property is
lineage continuity. Let $v_I$, $v_M$, and $v_B$ identify the attributable
identity version, memory snapshot ancestry, and body revision. A pure runtime
migration from $t$ to $t+1$ requires

\begin{equation}
\begin{aligned}
  v_I(t+1) &= v_I(t),\\
  v_M(t) &\preceq v_M(t+1),\\
  v_B(t+1) &= v_B(t),
\end{aligned}
\end{equation}

unless the migration transaction also contains a separately authorized memory
repair, body evolution, or identity update. Here $\preceq$ denotes an
auditable continuation rather than equality: checkpoint metadata, schema
migration, or new experience may extend memory without resetting its ancestry.
Schema conversion can preserve $B$ when that revision already implements the
required migrator; changing the migrator's code is a separate body evolution.

This account is causal and administrative rather than purely behavioral.
Different models may produce different wording, competence, or style while
executing the same persistent substrate. Behavioral identity fidelity is an
important empirical property, but it is not a reliable storage identifier and
cannot by itself prevent a copied process from claiming sole authority.

\subsection{Lifecycle operations}

Table~\ref{tab:operations} separates operations that are often collapsed into
``resume'' or ``fork.''

\begin{table}[t]
\centering
\small
\caption{Persistent-agent lifecycle operations.}
\label{tab:operations}
\begin{tabular}{@{}p{0.14\linewidth}p{0.38\linewidth}p{0.39\linewidth}@{}}
\toprule
Operation & Change & Continuity consequence \\
\midrule
Restart & Recreate a process in the same logical environment & System continuity retained; no binding migration \\
Resume & Reopen a session or queued task & System continuity retained if the same substrate is reloaded \\
Migrate & Replace one or more elements of $\executionenv$ or $\interactions$ & Authorized system continuity after invariant verification \\
Evolve & Change $B$ through its governed version lineage & Substrate lineage retained; new body revision \\
Identity update & Change $I$ under its declared authority rules & Same lineage, new identity version \\
Replicate & Instantiate the same checkpoint concurrently & Shared lineage; unique authority only if coordinated \\
Descend / fork & Create an independently governed substrate & New identity and private-state boundary \\
\bottomrule
\end{tabular}
\end{table}

Replication exposes why a stable name or UUID is insufficient. If a directory
is copied to two servers, both copies contain the same identifier. Without an
authority lease, fencing epoch, or declared multi-embodiment policy, the copies
cannot both safely behave as the unique continuation. Agent continuity requires
identity, attributable lineage, and transition authority; a stable label alone
is insufficient.

\section{Architecture}

\subsection{Provider-neutral binding}

Runtime independence requires the persistent body to depend on semantic
contracts rather than infrastructure brands. The body may include adapters,
but application logic should receive normalized operations and opaque provider
identities. A chat event, runtime result, repository revision, review unit, or
service descriptor must remain meaningful without exposing a Telegram update,
Codex transcript, Git branch, GitHub pull request, or launchd property list to
the portable core.

This separates capability from attachment. The body owns the capability and
policy for communication; an interaction surface supplies the current account,
channel, session, and provider-native identifiers. Rebinding Telegram to Slack
changes $\interactions$, not $I$, while removing every chat surface leaves an
agent that still exists but is temporarily unreachable through chat.

Each provider declares a kind, contract version, stable identity, and supported
capabilities. Before a side effect, the application verifies both the task's
requirements and the selected provider's grants. A provider swap may reduce
capability -- for example, a local review provider may not publish remotely --
without changing who the agent is. The missing capability must be explicit and
must fail closed rather than silently selecting another provider.

\subsection{Storage and custody boundaries}

The architecture separates three physical ownership areas:

\begin{itemize}
  \item the \textbf{software body}, reviewed and versioned;
  \item \textbf{private state}, including installed identity, memory,
  configuration, credentials, queues, schedules, sessions, and provider cursors;
  and
  \item retained \textbf{artifacts}, including logs and task, evolution, and
  learning evidence, which may have a different retention policy.
\end{itemize}

The logical substrate $(I,M,B)$ can therefore be transported through different
mechanisms. Public code may move through version control, private memory through
encrypted backup, and secrets through a target-host credential store. A
migration manifest records exact versions and hashes without requiring every
private byte to enter the body repository.

Human custody remains explicit. The agent can participate in checkpointing,
validation, provider selection, and health checks, but deployment authority,
secret release, identity update, and promotion policy remain externally
governed unless a custodian deliberately delegates them.

\subsection{Execution reliability across replacement}

A provider-independent state machine must survive the provider it invokes.
Before an external effect, the application records intent with an idempotency
key and a daemon fencing epoch. A stale process cannot claim or finalize work.
After ambiguous failure, a provider with reconciliation can prove whether the
effect occurred; a provider with idempotent delivery can safely replay it. A
legacy provider that offers neither must fail closed rather than risk a
duplicate message or action. Queue state and task identity likewise remain in
the persistent substrate, not in a transient harness transcript.

These rules separate \emph{session continuity} from \emph{agent continuity}. A
new harness session is expected after migration. The agent reconstructs its
context from identity, memory, body, and durable work records instead of
treating one provider-native conversation as its mind. A transcript is raw
interaction evidence, not automatically memory; selected facts, commitments,
relationships, and pending work enter $M$ through explicit retention and
workflow policies.

\section{Authorized Migration Protocol}

\begin{figure}[t]
  \centering
  \includegraphics[width=\linewidth]{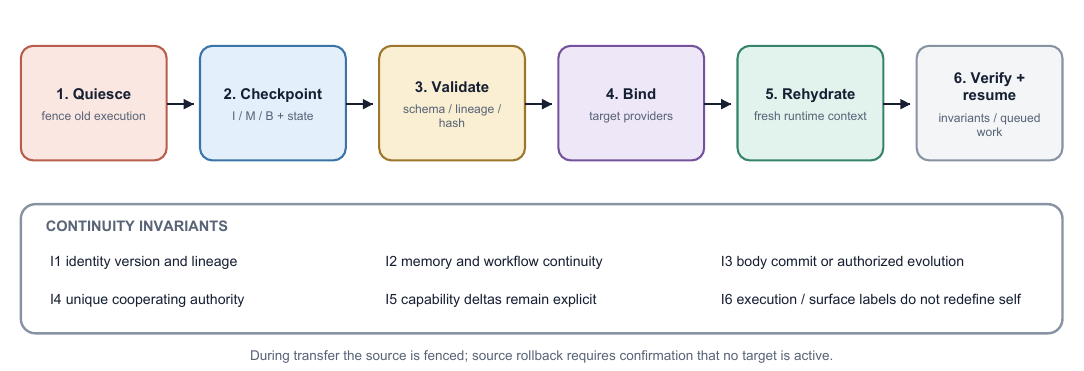}
  \caption{Authorized migration. The old execution is fenced before a
  checkpoint is captured. The target becomes authoritative only after binding,
  rehydration, and system-continuity verification succeed.}
  \label{fig:migration-protocol}
\end{figure}

Let $\mu:(\executionenv_a,\interactions_a)\rightarrow
(\executionenv_b,\interactions_b)$ replace any nonempty subset of reasoner,
harness, host, or interaction surface. Figure~\ref{fig:migration-protocol}
shows the six migration phases.

\begin{enumerate}
  \item \textbf{Quiesce and fence.} Stop admitting new work, allow or cancel
  bounded in-flight work, and advance an authority epoch so executions and
  providers that honor the governed binding reject work under the stale epoch.
  \item \textbf{Checkpoint.} Capture the identity version, body revision,
  memory and workflow-state versions, pending work, provider cursors, and
  artifact references. Secrets may be referenced rather than exported.
  \item \textbf{Validate.} Verify schemas, hashes, declared lineage,
  supported-state versions, and target capability requirements before mutating
  the target.
  \item \textbf{Bind.} Resolve target providers and credentials through the
  body contracts. Provider-native identifiers remain target-local.
  \item \textbf{Rehydrate.} Install or convert private state without admitting
  work on a partial checkpoint, load body and self as separate startup inputs,
  and create fresh harness and interaction sessions.
  \item \textbf{Verify and resume.} Run health and mechanical-continuity checks,
  acquire the new authority epoch, reconcile ambiguous effects, and resume the
  same durable tasks.
\end{enumerate}

The source is quiescent and fenced during transfer. Before target activation,
a failed conversion must restore modified target files or leave the target
inactive pending recovery. Resuming the source requires an explicit rollback
decision establishing that no target is active.
The reference implementation provides an operator-confirmed export-cancellation
path for that case. After target activation, the source must remain fenced even
if it reconnects. This defines a single governed promotion point, not a
distributed consensus or automatic cross-host atomic-commit protocol.

The protocol specifies six invariants that a valid migration must satisfy;
Table~\ref{tab:invariant-evidence} maps their implemented checks and observed coverage:

\begin{description}[leftmargin=0pt,labelwidth=0pt,beginpenalty=10000]
  \item[I1 -- Identity and lineage.] Migration preserves installed identity
  version and attributable lineage; updates require declared governance.
  \item[I2 -- Memory.] Memory and continuity-relevant workflow state extend or
  validly migrate their recorded ancestry; they are not silently reset.
  \item[I3 -- Body.] The target executes the same body revision unless a
  separately governed evolution is declared.
  \item[I4 -- Authority.] Within the governed deployment boundary, at most one
  cooperating execution can hold continuation authority and produce
  authoritative external effects for the migrated instance.
  \item[I5 -- Capability.] Environmental capability deltas are visible and do
  not masquerade as identity changes.
  \item[I6 -- Self-description.] Execution, process, interaction-surface, and
  deployment labels remain environment metadata and do not overwrite the
  installed identity.
\end{description}

\paragraph{Authority and failure assumptions.}
I4 assumes trusted custodians authorize promotion and rollback, participating
executions honor the governed authority records, and effects pass through
interfaces that enforce fencing and any required idempotency or reconciliation.
Crashes, restarts and delayed replies may occur within this boundary. The
implementation's local guards reject stale callers; they do not imply that
Slack, Telegram or another remote service checks Enoch epochs. A detached or
malicious copy with independently usable credentials is outside this boundary.
Enforcing authority against such copies requires provider-side fencing,
credential revocation or an external authority service. Section~\ref{sec:authority-realization}
describes the reference implementation's local and cross-host responsibilities.

I6 is easy to overlook. An agent body, package, bot account, host, and model
may each have a name. A fresh session may report one of those labels when asked
who it is, even though the installed identity says otherwise. The architecture
therefore loads identity and body as separate, explicitly labeled inputs and
treats behavioral self-representation as a post-migration test rather than an
assumption.

\input{implementation}

\input{evaluation}

\section{Discussion}

\subsection{Agent execution is not the longitudinal agent}

The model--harness view and the persistent-substrate view operate at different
time scales. At one instant, model and harness strongly determine what the
system can do. Across time, they are execution resources. Our claim is not that
models are interchangeable in quality or personality. It is that a change in
reasoning machinery need not, by architectural definition, break the authorized
system lineage when the continuity-bearing substrate and authority remain.

This distinction resembles a logical actor whose location and worker process
can change while its address and state persist~\cite{agha1986actors}. Persistent
agents add richer problems: identity representations, autobiographical memory,
executable self-modification, human relationships, model-dependent behavior,
and governance over copying. Distributed-systems mechanisms are therefore
necessary but not sufficient.

\subsection{Copying memory does not copy authority}

\emph{State similarity alone does not establish continuity; attributable
lineage and continuation authority are also required.} Two copies can hold
identical memories and return the same identity answers, yet only the
authorized successor may resume as the sole continuation. The migration
protocol makes that distinction explicit through a single promotion point.
A copied checkpoint may instead serve as a replica or descendant under its
declared authority policy.
Future multi-embodiment agents will require an explicit policy for concurrent
instances, reconciliation, and responsibility attribution.

\subsection{Safety and governance}

Runtime portability can improve user custody and reduce provider lock-in, but
it can also move an agent into a less capable security boundary. Migration must
therefore preserve or deliberately revise permissions, secret handling,
logging, and approval policy. Capability loss should fail closed. Capability
gain should require authorization rather than follow automatically from the
target server. Identity continuity does not imply that every remembered goal
or requested action is safe or authorized.

\section{Related Work}

PAI-Bench~\cite{zhao2026pai} evaluates fidelity to a versioned identity
contract with scoring oracles outside the target process. Its campaigns
compare independently initialized target configurations rather than following
one agent through runtime migration. RIPA addresses state lineage and
continuation authority across actual rebinding; PAI supplies a complementary
behavioral evaluation protocol. Our narrow recall witnesses do not constitute
a full PAI-Bench evaluation or inherit its experimental results.

Language-agent architectures describe combinations of models, memory, tools,
actions, and control flow. CoALA provides a cognitive architecture with modular
memory and action spaces~\cite{sumers2023coala}; AutoGen provides programmable
multi-agent interaction among models, humans, and tools~\cite{wu2023autogen}.
These frameworks explain agent execution and composition. Our focus is the
longitudinal boundary when the executing composition itself changes.

Distributed-systems research already separates logical entities and durable
computation from physical execution. Actor systems provide stable logical
addresses, and Orleans virtual actors persist beyond any in-memory activation
or particular server~\cite{agha1986actors,bernstein2014orleans}.
Checkpoint/restart captures consistent process state and can resume it on a
different host~\cite{laadan2007checkpoint}; mobile-code work classifies the
relocation of code, state, and execution~\cite{fuggetta1998mobility}; and
durable workflows persist and replay progress across process
failures~\cite{burckhardt2021durable}. These systems establish logical
addressing, host relocation, state recovery, and durable progress. Our
contribution is to place architectural identity, private memory, a versioned
software body, attributable lineage, human custody, behavioral-fidelity
considerations, and unique continuation authority in one AI-agent lifecycle
model.

Persistent-memory systems support continuity across interactions: Generative
Agents combines memory, reflection, and planning~\cite{park2023generative},
while Voyager accumulates executable skills~\cite{wang2023voyager}. Identity
work provides stability metrics~\cite{perrier2025identity} and redundant
anchors separated from memory~\cite{menon2026persistent}. We treat identity,
memory, and body as jointly continuity-bearing and add migration authority and
execution/interaction rebinding. Portable Agent Memory specifies verified
transfer and rehydration of structured memory across heterogeneous
agents~\cite{ravindran2026portable}, including identity preferences, provenance
and scoped access. RIPA's portability unit additionally binds the installed
identity and memory to a versioned executable body and governs the authority
to continue that agent under new execution and interaction bindings.

Most directly adjacent, the Continuity Kernel defines authorized state-head
lineage and formalizes atomic activation, writer handoff, schema migration, and
restoration~\cite{he2026continuity}. Its typed agent state extends beyond memory;
both works therefore treat authority and lineage as essential. CK specifies
the transactional admission of a successor state. RIPA specifies the
$(I,M,B)$ agent boundary and the continuity obligations when reasoner, harness,
host or interaction bindings change. The contracts are complementary: a
CK-style activation layer could realize RIPA's authority requirement. Our
Enoch experiments exercise deployment rebinding, without evaluating or
inheriting CK's formal guarantees.

Self-evolving systems make implementation resources mutable: MOSS governs
source rewriting~\cite{cai2026moss}; Autogenesis versions agent resources under
a propose--assess--commit loop~\cite{zhang2026autogenesis}; SemaClaw explores a
persistent personal-agent harness~\cite{zhu2026semaclaw}; and Agent libOS adds
process lineage, capabilities, and approvals~\cite{zhang2026agentlibos}. Our
closest precursor is the agent-owned software body~\cite{zhao2026body}. We
generalize it to an explicit persistent/execution/interaction boundary and
whole-agent migration semantics.

\section{Limitations}
\label{sec:limitations}

\paragraph{Scope and external validity.}
All studies use one evolving Enoch body lineage. The operational observations,
single cross-host task and B2 repetitions are complementary cases, not a
controlled all-axis matrix. Claude changes both model and harness; Muse
changes several deployment components. Neither isolates a causal benefit of
one architecture axis. The synthetic tasks are simple, tool-free model
requests orchestrated by explicit drivers. They do not demonstrate migration
of arbitrary repository worktrees, active inference processes or provider-native
sessions. A single native completion in the observed trace is not a general
exactly-once guarantee for external effects.

\paragraph{Measurement and reproducibility.}
Source-state and artifact checks are more direct evidence than self-reported
identity. Private operational data, shared Muse conversation context,
operator-mediated execution and unavailable backend attestation constrain
behavioral interpretation and independent replication. The fault sample uses
one prescribed cut while waiting for a reply. The missing-record supplement,
read-only verifier adaptation and stale-helper exit-evidence gap are disclosed
in Section~\ref{sec:evaluation}. We do not estimate statistical reliability,
operator burden, cost or model performance. Provider conformance detects
interface violations but does not prove semantic equivalence or security.

\paragraph{Authority and agent identity.}
The authority and failure assumptions in Section~4 bound I4; the experiments
do not establish distributed mutual exclusion or provider-side rejection of
detached copies. The
$(I,M,B)$ decomposition is functional, not a complete identity theory;
coordinated multi-embodiment and divergent-memory merging remain open.

\paragraph{Further evaluation.}
A broader study should separate model, harness, host and surface changes;
use independent reasoning sessions and evaluator-private behavioral probes;
exercise richer file-producing tasks and pending notifications; and inject
failures before and after promotion and external actions. Mechanical
continuity and behavioral identity fidelity must remain separate outcomes:
an authorized successor may behave differently, while an unauthorized copy may
answer identity questions correctly.

\section{Conclusion}

A persistent agent need not be the model, harness, or server executing it, nor
the chat carrying its messages. The first three form an execution substrate;
the latter is an interaction surface. Identity, memory, and a versioned
executable body form a continuity-bearing substrate whose replacement bindings
can be governed as auditable lifecycle operations.

Enoch realizes this boundary through separate body/identity loading, durable
private state, five provider contracts and guarded transitions. Operational
Claude, host and chat-surface substitutions preserve established instance
state; bounded cross-host and Muse studies continue checkpointed work, and a
planned Muse interruption recovers with stale-reply rejection. Together,
these cases demonstrate how an explicit agent boundary and governed migration
can preserve authorized system continuity across the tested deployments and
bounded workflows.

\clearpage
{\scriptsize
\bibliographystyle{unsrt}
\bibliography{references}
}

\end{document}

%% file: implementation.tex
\section{Reference Implementation}

\href{https://github.com/our-ark/enoch}{Enoch} is a reusable open-source agent
body, distinct from each installed agent's identity and private state. We
report three frozen body revisions rather than attributing all evidence to
one release (Table~\ref{tab:snapshots}). A body revision is held fixed within
each migration experiment; the later studies use subsequent body releases.

\begin{table}[t]
\centering\small
\caption{Evidence snapshots. Hashes are abbreviated here; full identifiers
and study summaries accompany the source as ancillary files. Test totals
belong to their respective snapshots and are not pooled.}
\label{tab:snapshots}
\begin{tabular}{@{}>{\raggedright\arraybackslash}p{0.13\linewidth}>{\raggedright\arraybackslash}p{0.23\linewidth}>{\raggedright\arraybackslash}p{0.54\linewidth}@{}}
\toprule
Snapshot & Frozen body & Evidence scope \\
\midrule
B0, 31 Aug. & \texttt{c8013ed249bc} & Original clean-room mechanism suite: 833 core and 92 provider/library tests \\
B1, 7 Sep. & \texttt{e40b28782f5c} & Operational migrations and cross-host task; 857 core and 106 provider/library tests in original and anonymized snapshots \\
B2, 19 Sep. & \texttt{66781e209962} & Muse round trips, controls and planned fault; adapter target \texttt{68e94efd9896}, with 23 adapter tests \\
\bottomrule
\end{tabular}
\end{table}

\subsection{Reusable body and installed agent instance}

Enoch loads \texttt{body.yaml} from the software body and a private
\texttt{self.json} from the installed agent. The former is a manifest of
body/package identity, mission, principles and repository lineage; the
repository supplies the executable code, prompts, tools and policies. The
latter stores the installed identity, including designation, relationships
and values, alongside separate lineage metadata. An installed agent is the
durable binding $\agentcore_t^{(a)}=(I_t^{(a)},M_t^{(a)},B_t^{(a)})$,
governed by continuation-authority metadata. Copying the body alone does not
copy the agent.

Body and identity reload separately for fresh runtime sessions. Memory,
configuration, queues, schedules, provider cursors and authority records live
in private state; retained evidence uses an artifact namespace. Versioned
schemas support read-only validation, backup before conversion, manifest-last
commit and restoration after caught conversion or validation failures. Host migration additionally
packages portable state and explicitly selected artifacts, fences the exported
source, and verifies the imported checkpoint before target activation.

\paragraph{Recovery boundary.}
Schema conversion backs up state, replaces individual files atomically, then
writes the manifest; failed validation triggers restoration. Import records an
\texttt{importing} marker, writes files and marks the completed import ready
for activation. Tests inject write and validation exceptions and check cleanup
or restoration. These mechanisms do not make the multi-file operation
crash-atomic: SIGKILL bypasses exception rollback, and automatic recovery from
every partial-write or interrupted-rollback state has not been demonstrated.
An incomplete import cannot be activated through the normal API. B2 transfers
already-current schemas; its worker interruption does not test conversion or
import crashes.

\subsection{Provider contracts and runtime adapters}

Five provider kinds isolate infrastructure dependencies
(Table~\ref{tab:providers}). The private instance configuration selects
providers without changing the body. Runtime providers own model identifiers
and reasoning settings, accept fresh sessions or task session keys, and return
normalized progress and completion records. Conformance tests exercise
contracts; live substitution cases supply separate execution evidence.

\begin{table}[t]
\centering\small
\caption{Provider contracts and implementations used across the snapshots.
Muse is an external adapter; its live evaluation is supervised.}
\label{tab:providers}
\begin{tabular}{@{}p{0.12\linewidth}p{0.29\linewidth}p{0.49\linewidth}@{}}
\toprule
Kind & Implementations & Portable responsibility \\
\midrule
Chat & Telegram, Slack; Muse adapter & Normalized events, messages, acknowledgements and cursors \\
Runtime & Codex, Claude Code; Muse adapter & Reasoning requests, sessions, progress, completion and cancellation \\
VCS & Git; branchless fixture & Revision inspection, capture, isolation and restoration \\
Review & GitHub, local; independent fixture & Review creation, inspection, stacking and landing \\
Service & launchd, systemd & Process installation, start, stop, inspection and diagnosis \\
\bottomrule
\end{tabular}
\end{table}

The open-source \href{https://github.com/our-ark/enoch-muse-runtime}{Muse adapter}
bridges Enoch's runtime contract to an asynchronous file
mailbox. Each request carries a fresh attempt nonce; a response is accepted
only when it matches that attempt. A separate chat adapter transports events
through Enoch's native daemon. The body continues to own installed identity,
memory, workflow state and authority; Muse supplies reasoning through the
external consumer. This is a different deployment binding, not a new installed
identity. In the reported experiments, a live Muse conversation serves the
mailbox through operator-mediated UI interaction.

The managed launcher registers providers through Enoch's public registry,
persists target bindings, and keeps each instance's mailbox separate. A
per-root process lock prevents duplicate managed launches; the child inherits
the lock so that a surviving child continues to exclude another launch after
its supervisor exits. Each attempt retains command, process, exit and log-hash
evidence. Recovery first establishes that the recorded worker has exited and
then invokes the native task-pause API with the retained checkpoint. It neither
rewrites the queue directly nor automatically retries the task. These adapter
mechanisms made supervised migration and failure observation concrete without
modifying the pinned B2 body.

\paragraph{Authority realization.}
\label{sec:authority-realization}
Three mechanisms have distinct scopes. The Muse per-root lock rejects a
duplicate managed launch before it can advance the epoch. Within an installed
root, Enoch stores an epoch token and generation in private state; guarded
effects check the token while holding the epoch lock. Across hosts, export
fences the source, and the imported manifest supplies a generation floor:
the target's next generation exceeds both its local value and that floor.
This transfers a recorded authority boundary, not a shared cross-host lock.
Source rollback removes the export fence only under an operator decision
that no target is active; it neither restores an older epoch nor independently
queries the target. Thus cross-host exclusivity depends on governed promotion
and custody, while local stale-process rejection is mechanically enforced.

\subsection{Mechanism evidence and its scope}

The B0 clean-room suite used CPython 3.12.13. B1's original and anonymized
snapshots each passed 963 tests, with no failures or skips; the anonymized
distribution was unpacked and tested separately. B2's 23 adapter tests include
native daemon and workflow paths with deterministic mailbox responses. They
test implementation mechanisms, not model ability, and are not additional
live migration samples. Existing suites cover identity/body separation,
schema rollback, provider selection, guarded task ownership, stale-token
rejection and notification recovery. The studies below test bounded uses of
these mechanisms across actual deployment bindings.

%% file: evaluation.tex
\section{Evaluation}
\label{sec:evaluation}

We ask three systems questions: (RQ1) which continuity-bearing state survives
actual deployment substitutions; (RQ2) whether an explicitly checkpointed task
can continue under a new binding; and (RQ3) whether a worker interruption can
be recovered while rejecting a stale reply. We distinguish operational
observations, a single cross-host task, and a frozen Muse study. They use
different instances, workloads and body revisions, so their outcomes are not
combined into a single success rate.

Table~\ref{tab:invariant-evidence} separates concrete mechanism assertions
from live observations. The ancillary \texttt{INVARIANTS.md} identifies the
pinned test functions and recorded verifier checks behind each row. This is a
coverage map of existing evidence, not a new experiment or proof of the full
contract.

\begin{table}[t]
\centering\small
\caption{Invariant-to-evidence map. B1 and B2 denote the frozen snapshots in
Table~\ref{tab:snapshots}. Mechanism tests and live cases provide partial,
distinct coverage; no row establishes the invariant for all deployments.}
\label{tab:invariant-evidence}
\begin{tabular}{@{}>{\raggedright\arraybackslash}p{0.07\linewidth}>{\raggedright\arraybackslash}p{0.28\linewidth}>{\raggedright\arraybackslash}p{0.31\linewidth}>{\raggedright\arraybackslash}p{0.25\linewidth}@{}}
\toprule
Invariant & Mechanism / assertion & Existing evidence & Coverage boundary \\
\midrule
I1 & Load installed identity and lineage; preserve exact identity record & Identity-loading / declared-lineage assertions; B1 hashes; B2 exact identity-record equality & No live governed identity-update or divergent-lineage trial \\
I2 & Retain memory ancestry, task ID and checkpoint; reject corrupt artifacts & B1 checkpoint replay; B2 memory/task checks; F preserves the paused checkpoint & Exception rollback tests do not establish crash-atomic conversion \\
I3 & Pin body revision and reject mismatched imports & Body-mismatch test; B1 and B2 recorded revision checks & No combined live migration and body evolution \\
I4 & Reject stale epochs and exported-source starts; correlate reply attempts & Stale-effect and duplicate-launch tests; B1/B2 fences; F rejects an old nonce & Nonce checks are attempt-local; no external-effect crash or detached-copy trial \\
I5 & Check provider grants before effects; report missing capability & Deterministic task test rejects missing workspace isolation with no provider effects & No targeted live capability-loss migration trial \\
I6 & Load self and body separately; keep installed designation & Self-summary fixture distinguishes self from body; B2 daemon identity recall & No conflicting-environment probe or broad behavioral-fidelity result \\
\bottomrule
\end{tabular}
\end{table}

\subsection{Operational migrations, including Claude Code}

The B1 observational case follows one existing installed agent with nonempty
private long-term memory. It moved from a laptop to a desktop and back while
retaining Codex and Slack, then replaced Codex with Claude Code on the laptop
while retaining Slack, and finally changed Slack to Telegram while retaining
Claude Code. Recorded identity and memory hashes, the body revision and
22 historical task IDs were preserved. Outbound and return host checkpoints
verified 76 and 77 portable files respectively; the former source remained
export-fenced after handoff. The successive recorded authority generations
were 8 through 14 across export, activation and service starts.

The joint model/harness transition requested
\texttt{gpt-5.6-sol}/high through Codex CLI 0.153.1 and then
\texttt{claude-opus-5}/high through Claude Code 2.1.258. Twelve live Claude
probe responses recorded the latter model identifier. This demonstrates an
actual second reasoning harness, but does not isolate a model-only effect or
establish equivalent inference budgets. These are recorded client/runtime
identifiers, not independent backend attestations. Telegram startup delivery
and session creation were observed; a user-message-to-agent-reply round trip
was not tested in that case. Cursors remain provider-scoped; this surface
change does not demonstrate translation of unread-message position or delivery
guarantees between Slack and Telegram.

These observations support continuity of an established installed instance
(RQ1), with no unfinished task or pending external effect injected. The records
are author-curated observations derived from private state. Ancillary files
provide a redacted summary rather than the private identity, memories or raw
conversations.

\subsection{Cross-host continuation of a checkpointed task}

A separate B1 synthetic case starts one two-stage task on a laptop and
continues it on a desktop. Both stages use Codex CLI 0.153.1, requested
\texttt{gpt-5.6-sol}/high, and Python 3.12.14. An external experiment worker
calls Enoch's existing workflow and migration APIs. The first model call
normalizes four records; the worker persists the result, a relative artifact
reference, its hash and the next-stage index, then pauses and exports the task.
The target imports nine portable files, verifies and consumes the retained
artifact, resumes the same task ID, and computes category totals of 28 and 28,
with grand total 56. Both model calls return the fixed-oracle result on their
first attempt. Task attempts advance from 1 to 2 and authority from 1 to 2 to 3;
the exported source refuses restart.

Identity, body, memory and checkpoint checks pass, and the event trace records
one execution of each stage. The synthetic long-term-memory fixture is empty;
the nonempty migrated work is the normalized-record artifact and workflow
descriptor. A deterministic offline replay uses the same native APIs and
rejects missing or corrupt continuation artifacts. The target's code-mode
helper warning is retained: neither stage requested tools. This case answers
RQ2 for an explicit safe checkpoint and bounded task.

\subsection{Frozen Codex--Muse--Codex study}

\paragraph{Protocol and execution.}
The B2 study is \emph{supervised}: an operator mediates the live Muse mailbox
interaction and coordinates execution. It freezes the workflow driver, inputs, body and adapter before
source outcomes. It schedules one qualification round trip Q, five ordinary
round trips R1--R5, five paired same-runtime controls C1--C5, and one planned
fault case F. The earlier development pilot required operator recovery and is
retained separately; it is excluded from this measured series. All started
cases in the frozen series are reported.

Each case has a fresh installed identity, three integer ledger entries, a
multiplier and distinct source/target memory markers. A real Codex call
recalls the supplied identity and source marker and computes the subtotal;
the driver stores the result and pauses the task through the native API.
Export fences the source. A fresh Muse-hosted root imports, activates and
verifies the bundle. A separate daemon conversation probes identity and
memory through the Muse chat and runtime providers. After that daemon stops,
the explicit workflow driver resumes the original task through Muse, obtains
the multiplied subtotal, writes the target marker and completes the task.
A return export fences the Muse root. A fresh Codex-hosted root imports and
verifies the return bundle, and a real Codex call recalls both markers and the
final result. These simple computations provide exact continuity witnesses.

Source and return use Python 3.13.14 and Codex CLI 0.153.1, configured as
\texttt{gpt-6-astra}/xhigh; Muse uses Python 3.12.3. The served Codex backend is
not independently attested, and Muse's backend model is unknown. The adapter's
target commit is \texttt{68e94efd9896}; Muse's local deployment commit is
different, but the checked source tree matches
\texttt{643dfa19e693f3d519edc7fd7712c256bdd196c7}. Version provenance therefore
uses tree equality, not a claim of identical local commit IDs.

Muse replies are produced from actual mailbox requests through a live operator
conversation. Coordination includes continuation prompts between cases.
Fresh installed roots do not imply independent or blinded Muse reasoning
sessions: the remote cases share the operator conversation. An author-run
verifier outside the target agent checks exported state and execution records.

\paragraph{Controls and acceptance checks.}
Each control matches its paired round trip's ledger, multiplier and canonical
name, but uses a separate identity ID and memory markers. Its source,
continuation and recall run as separate Codex processes in one host/root,
using native pause/resume without export or the cross-host daemon probe.
Controls establish that the same bounded workflow also completes without
migration; they do not isolate host, runtime and interaction-surface effects.

The round-trip verifier checks 28 conditions spanning model JSON witnesses,
exact identity bytes, memory preservation and extension, body revision, task
identity, checkpoint consumption, one completion record/event, source-event
prefix preservation, empty outstanding work, authority progression, distinct
migration IDs, native import verification, both source fences, Muse binding
metadata, nonce-correlated worker replies, daemon delivery and observed
process exits. Each control has 13 applicable checks. An additional audit
checks 18 source/driver/log-provenance and no-unplanned-repair conditions per
ordinary round trip. Check counts describe assertions, not independent
statistical samples or exhaustive validation of all six invariants.

\begin{table}[t]
\centering\small
\caption{B2 study outcomes. Qualification, ordinary repetitions, controls and
the planned fault are separate groups. All were supervised. A fault-case
completion is not an ordinary first-attempt success.}
\label{tab:muse-results}
\begin{tabular}{@{}p{0.19\linewidth}p{0.10\linewidth}p{0.25\linewidth}p{0.35\linewidth}@{}}
\toprule
Group & Cases & Observed outcome & Verification per case \\
\midrule
Qualification Q & 1 & First-attempt pass & 28 continuity/evidence checks \\
Round trips R1--R5 & 5 & 5 first-attempt passes & 28 continuity/evidence + 18 provenance/no-repair checks \\
Controls C1--C5 & 5 & 5 first-attempt passes & 13 applicable workflow checks \\
Planned fault F & 1 & First worker killed; native recovery completed & 28 continuity/evidence + 29 provenance/recovery checks \\
\bottomrule
\end{tabular}
\end{table}

\paragraph{Results.}
Q and all five ordinary round trips pass on their first attempts, and all five
controls pass (Table~\ref{tab:muse-results}). There are no model-answer retries
or unplanned state repairs in this supervised series. All seven migrating
source roots and seven Muse return-export
roots reject post-export execution without advancing the authority epoch.
The observed continuations retain identity, source memory and the pinned body,
extend memory with the target marker, and finish the same task (RQ1--RQ2).
The arithmetic inputs and per-case outcomes are retained in the ancillary
study summary.

\subsection{Planned interruption and recovery}

\begin{figure}[t]
\centering
\includegraphics[width=\linewidth]{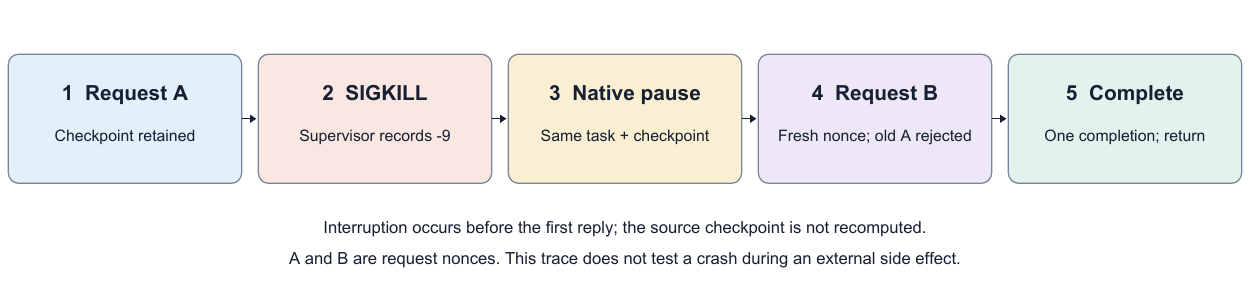}
\caption{Recorded recovery path in F. A and B denote distinct request-attempt
nonces, not authority epochs. The old-nonce rejection and successful B reply
are both retained. Exactly one native task completion is observed in this trace.}
\label{fig:recovery}
\end{figure}

F kills the first Muse worker after request publication and before a reply.
Its supervisor records return code $-9$ (SIGKILL). Native recovery pauses the
same task with its full stage-one checkpoint unchanged. A subsequent worker
uses a fresh attempt nonce, completes the task once, and preserves that state
on return to Codex. Submission of the old nonce produces rejection stderr
(Figure~\ref{fig:recovery}). This answers RQ3 at the selected wait-for-reply
cut; it does not test a crash during an external side effect.

The first returned fault package omitted the saved native recovery record and
logs. A supplement supplied those existing bytes, whose hashes matched the
original status manifest; no execution was rerun. The read-only verifier was
also adapted to F's two worker directories and a plain-text stderr file with a
JSON suffix. The original failed analysis log and both parser hashes are
retained; experimental criteria and driver were unchanged. All 28 continuity
and 29 recovery/provenance checks then pass. The stale-reply helper's reported
exit code 3 is only an operator observation, with no separate machine exit
record; the rejection claim rests on retained stderr. Worker termination,
recovery and successful continuation have captured process evidence.

\subsection{Evidence availability and interpretation}

Ancillary files contain the redacted operational record, the earlier synthetic
task record, B2 inputs and outcome summaries, full revision identifiers, and
a claim-to-evidence map. Enoch's body and the Muse adapter are public. The
B2 Muse experiment archive is pinned at commit
\texttt{084e1b1c6a54af749fb28a20e4745025badc0899} in the
\href{https://github.com/our-ark/enoch-muse-runtime}{adapter repository}.
Its Apache-2.0 release covers code, protocol, raw records and migration
bundles. An offline verifier checks hashes, arithmetic and bounded state
consistency without model calls; it does not attest the original hosts.
New live runs require Codex/Muse access and an operator. B1's private
operational identity, memory and conversations remain withheld.

The strongest common result is mechanical continuity under the tested
bindings and bounded workflows. Identity recall corroborates state loading,
but is weaker evidence of broad behavioral fidelity, especially with shared
Muse conversation context. Muse reports placeholder zero token counts, treated
as missing; wall times include operator/transport delay and overlapping local
batches. Section~\ref{sec:limitations} consolidates the scope and measurement
limits of these studies.